\documentclass[
 preprint,
 amsmath,
 showkeys
]{revtex4-1}
\usepackage{graphicx}
\usepackage{overpic}
\usepackage{multirow}
\usepackage{appendix}

\begin{document}

\title{Radiation Reaction of Betatron Oscillation in Plasma Wakefield Accelerators}
\author{Ming Zeng}
\email[Author of correspondence: ]{zengming@ihep.ac.cn}
\affiliation{Institute of High Engergy Physics, Chinese Academy of Sciences, 100049 Beijing, China}
\author{Keita Seto}
\affiliation{Extreme Light Infrastructure - Nuclear Physics, Horia Hulubei National Institute for Physics and Nuclear Engineering, 077125 Magurele, Romania}
\today

\begin{abstract}
A classical model of radiation reaction for the betatron oscillation of an electron in a plasma wakefield accelerator is presented. The maximum energy of the electron due to the longitudinal radiation reaction is found, and the betatron oscillation damping due to both the longitudinal and transverse radiation reaction effects is analyzed. Both theoretical and numerical solutions are shown with good agreements. The regime that the quantum radiation takes effect is also discussed. This model is important for designing future plasma based super accelerators or colliders.
\end{abstract}

\maketitle
\section{Introduction}
Plasma Wakefield Accelerators (PWA), driven by either short laser pulses or charged particle beams~\cite{TTajimaPRL1979, PChenPRL1985, CJoshiPOP2020}, have the electric (E) and magnetic (M) fields $E_0 \sim 96\sqrt{n_p\left[\rm cm^{-3}\right]}\ \left(\rm V/m\right)$ and $B_0 \sim 320 \sqrt{n_p\left[\rm 10^{18}\ cm^{-3}\right]}\ \left(\rm T\right)$ which are commonly more than 2 orders of magnitude larger than the EM-fields in conventional accelerators~\cite{EEsareyRMP2009}. With the help of powerful wakefield drivers, people are developing compact high-energy accelerators around the world~\cite{MLitosNature2014, RAssmannPPCF2014, CALindstroemPRL2021, AJGonsalvesPRL2019, WTWangPRL2016}.

In a PWA, the transverse focusing force of the beam, due to the radial E-field and azimuthal B-field of the wake, exists along the whole accelerator. As a consequence, the normalized emittance of the beam may grow due to the betatron oscillation (BO) and finally saturate~\cite{XLXuPRL2014}, with the saturation normalized emittance commonly ranging from 0.1 to 100 $\rm mm\cdot mrad$, depending on the injection mechanism, e.g.\ self-injection, density gradient injection, optical injection, ionization injection and so on~\cite{SCordeNatComm2013, GWittigPRAB2015, GGolovinPRL2018, APakPRL2010, CMcGuffeyPRL2010, BHiddingPRL2012, LLYuPRL2014, MZengPRL2015, MZengPOP2016, MZengNJP2020}. 

Radiation reaction (RR) is the friction on an electron due to EM radiation of itself. The RR effect has been studied for laser-plasma interactions, typically on energetic electrons colliding with high-intensity laser pulses. It is predicted that more than 35\% of laser energy is converted to EM radiation at a laser radiance of $I \lambda^2  > 10^{22}\ \mathrm{W \mu m^2/cm^2}$ in a duration of 20 fs~\cite{Zhidkov2002}. The quantum correction in RR has been confirmed by recent experiments of the head-on collision between energetic electron bunches and high-intensity laser pulses~\cite{Poder2018, Cole2018}. These researches focused on an extreme condition of highly intense EM-field in a short period of the fs-order to study the elementary process of a radiating electron. On the other hand, the long-time accumulation of the RR effect has also been studied, e.g.\ for runaway electrons in Tokamak to engineer good plasma conditions of magnetic confinement plasma~\cite{Matsuyama2017}.

For a regular PWA length of the order of 0.1 m or shorter, RR is negligible because the RR force is several orders of magnitude smaller than the acceleration and focusing forces of the wake. However, the  magnitude of the RR force becomes significant at high beam energies. Moreover, the RR force is a damping force and its effect accumulates, which may finally become noticeable at long acceleration distances. Such effects are crucial for developing future super plasma wakefield accelerators and colliders which have acceleration distances of the order of 10 to $10^3$ meters ~\cite{WLeemansPT2009}.

In this work, we show a classical model of the RR effects for the single electron BO in a PWA. The theory is reinforced by numerical results obtained by the code PTracker~\cite{PTracker}, which uses 4-th order Runge-Kutta Method to solve the simplified equations of motion of an electron in a PWA. This paper is organized as follows. Section~\ref{sec:BO} gives the general form of BO of an electron in a PWA. Section~\ref{sec:x1_gamma} shows the BO amplitude scaling during acceleration or deceleration processes. Section~\ref{sec:RR} presents a classical model of RR during BO, and Section~\ref{sec:CQL} discusses the limit between classical and quantum RRs. 

\section{\label{sec:BO}Betatron Oscillations without Radiation Reaction}
For a single electron trapped (with relativistic longitudinal velocities) in a blowout plasma wakefield and moving in the $z$-$x$ plane, the focusing force (in $x$ direction) and accelerating force (in $z$ direction) near the center of the blowout bubble are~\cite{WLuPOP2006}
\begin{eqnarray}
	f^{\rm ext}_x&=&-\frac{1}{2}x, \label{eq:f_ext_x}\\
    f^{\rm ext}_z&=&-\frac{1}{2}\zeta, \label{eq:f_ext_z}
\end{eqnarray}
where $\zeta=z-\beta_w t$ is the wake co-moving frame with $\zeta=0$ and $x=0$ being the center of the blowout bubble, $\beta_w=\sqrt{1-1/\gamma_w^2}$ is the wake phase velocity normalized to speed of light in vacuum $c$, and $\gamma_w \gg 1$ is the Lorentz factor of the wake. Note we have adopted the normalized units where length is normalized to $k_p^{-1}$, time is normalized to $(ck_p)^{-1}$, momentum is normalized to $m_ec$, energy and work are normalized to $m_e c^2$, force is normalized to $m_ec^2k_p$, with $k_p=\sqrt{\mu_0e^2 n_p/m_e}$ being the wavenumber of plasma wake, $\mu_0$ being the vacuum permeability, $e$ being the elementary charge, $n_p$ being the plasma density and $m_e$ being the electron mass. Without RR, the equation of motion is
\begin{eqnarray}
	\dot{\vec{p}}&=&\vec{f}^{\rm ext},\label{eq:EOM_no_rr}
\end{eqnarray}
with
\begin{eqnarray}
	p_x&=&\gamma \dot{x},\\
    p_z&=&\gamma \left(\dot{\zeta} + \beta_w\right),\\
    \gamma^2 &=&1+p_x^2+p_z^2, \label{eq:gamma_def}
\end{eqnarray}
where a dot on top means time derivative, $\vec{p}$ is the normalized momentum and $\gamma$ is the Lorentz factor of the electron.

Assume $\left|x\right|$ is a small quantity. Then the trajectory in the wake co-moving frame features a figure ``8'' with the frequencies in $x$-axis being $\omega_\beta$ and in $z$-axis being $2\omega_\beta$ where $\omega_\beta=1/\sqrt{2\gamma}$ is the normalized BO frequency~\cite{EEsareyPRE2002}
\begin{eqnarray}
	x &=& x_1 \sin\omega_\beta t, \label{eq:betatron_x}\\
    \zeta &=& \zeta_0 - \zeta_1 \sin2\omega_\beta t, \label{eq:zeta}
\end{eqnarray}
where subscriptions 0 / 1 indicate ``slow'' / ``fast'' components with characteristic time scale much longer than / comparable to $\omega_\beta^{-1}$. The normalized velocities are
\begin{eqnarray}
	\beta_x &=& \dot{x} = x_1 \omega_\beta \cos\omega_\beta t, \label{eq:beta_x}\\
    \beta_z &=& \beta_w + \dot{\zeta} = \beta_{z0} - 2\zeta_1 \omega_\beta \cos2\omega_\beta t,
\end{eqnarray}
where $\beta_{z0}=\beta_w +\dot{\zeta_0}$ is $\beta_z$ averaged in the $\omega_\beta^{-1}$ time scale. Thus
\begin{eqnarray}
	\gamma^{-2} = 1-\beta_x^2-\beta_z^2 =\gamma_{z0}^{-2} - \frac{x_1^2\omega_\beta^2}{2} + \left(4\beta_{z0}\zeta_1\omega_\beta - \frac{x_1^2\omega_\beta^2}{2}\right)\cos2\omega_\beta t, \label{eq:gamma_expression1}
\end{eqnarray}
where high order terms are neglected, and
\begin{eqnarray}
    \gamma_{z0}=\frac{1}{\sqrt{1-\beta_{z0}^2}}. \label{eq:gamma_z0_beta_z0}
\end{eqnarray}

Meanwhile,
\begin{equation}
    \begin{aligned}
	    \dot{\gamma} &= \beta_x f^{\rm ext}_x + \beta_z f^{\rm ext}_z = -\frac{x\beta_x}{2} - \frac{\zeta\beta_z}{2}\\
	    &= -\frac{\zeta_0 \beta_{z0}}{2} + \zeta_0\zeta_1\omega_\beta\cos2\omega_\beta t + \left(\frac{\beta_{z0}\zeta_1}{2} - \frac{x_1^2\omega_\beta}{4}\right)\sin2\omega_\beta t. \label{eq:gamma_dot0}
	\end{aligned}
\end{equation}
Commonly in a plasma wakefield $\left|\zeta_0\right| \lesssim 1$. Note $\omega_\beta \ll 1$, the second term is negligible, thus
\begin{eqnarray}
	\gamma = \gamma_0 + \left(\frac{x_1^2}{8} - \frac{\beta_{z0}\zeta_1}{4\omega_\beta}\right)\cos2\omega_\beta t, \label{eq:gamma_expression2}
\end{eqnarray}
where
\begin{eqnarray}
	\left.\gamma_0\right|_{t_0}^{t_1} = - \frac{1}{2} \int_{t_0}^{t_1} \zeta_0 \beta_{z0} dt \label{eq:gamma0}
\end{eqnarray}
is $\gamma$ averaged in the $\omega_\beta^{-1}$ time scale.

We apply Eq.~(\ref{eq:gamma_expression2}) to Eq.~(\ref{eq:gamma_expression1}) and multiply both sides by $\gamma_0^2$. By collecting the ``slow'' varying parts one obtains
\begin{eqnarray}
	\gamma_{z0} = \frac{\gamma_0}{\sqrt{1+\frac{x_1^2\omega_\beta^2\gamma_0^2}{2}}} = \frac{\gamma_0}{\sqrt{1+\frac{x_1^2\gamma_0}{4}}} \label{eq:gamma_z0}
\end{eqnarray}
which can be applied to Eq.~(\ref{eq:gamma_z0_beta_z0}) for the averaged longitudinal velocity $\beta_{z0}$. Note
\begin{eqnarray}
    \omega_\beta \approx \frac{1}{\sqrt{2\gamma_0}} \label{eq:omega_beta}
\end{eqnarray}
because $\gamma = \gamma_0 + \mathcal{O} \left(x_1^2\right)$. $\gamma_{z0}$ is regarded as the phase-locking Lorentz factor. If $\gamma_w = \gamma_{z0}$ and $\zeta_0=0$ initially, $\zeta_0$ will remain 0 and there will not be net acceleration or deceleration. By collecting the ``fast'' oscillation parts one finds the longitudinal oscillation amplitude
\begin{eqnarray}
	\zeta_1 = \frac{1-2\omega_\beta^2\gamma_0^3}{1-8\omega_\beta^2\gamma_0^3}\cdot\frac{\omega_\beta x_1^2}{2\beta_{z0}} \approx \frac{\omega_\beta x_1^2}{8} \approx \frac{x_1^2}{8\sqrt{2\gamma_0}}. \label{eq:zeta1}
\end{eqnarray}

\begin{figure}
    \centering
    \begin{overpic}[width=0.49\textwidth]{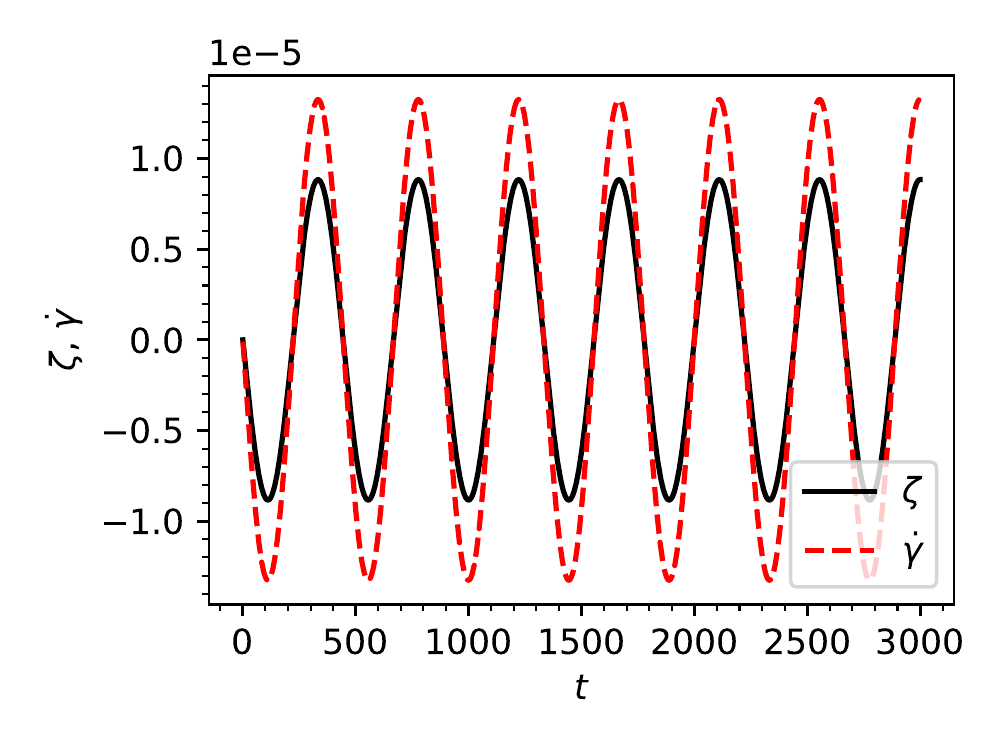}
        \put(23,60){(a)}
    \end{overpic}
    \begin{overpic}[width=0.49\textwidth]{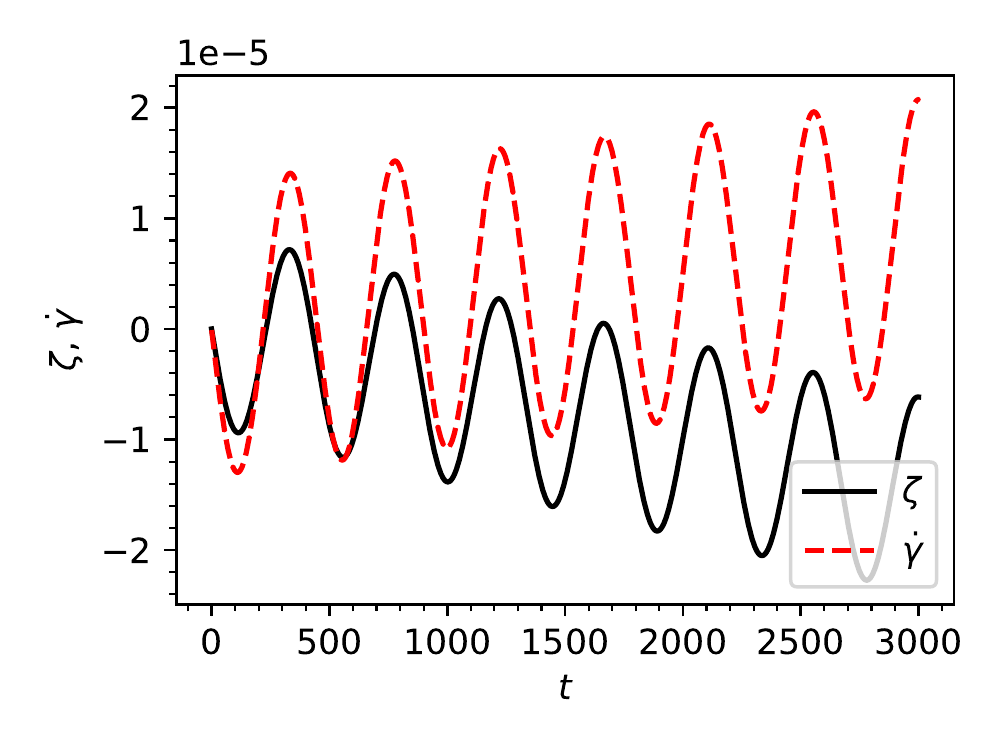}
        \put(20,60){(b)}
    \end{overpic}
    \caption{Numerical results of BO by the code PTracker with RR turned off, showing $\zeta$ and $\dot{\gamma}$ vs.\ $t$. (a) The parameters at $t=0$ are $\gamma_0=10^4$, $x_1=0.1$, $\zeta_0=0$ and $\gamma_w=1961.16=\gamma_{z0}$ according to Eq.~(\ref{eq:gamma_z0}). (b) The parameters are similar to (a), but with $\gamma_w=2000>\gamma_{z0}$ which makes the electron phase-shift towards the $-\zeta$ direction.}
    \label{fig:no_RR}
\end{figure}

We show two numerical results obtained by PTracker (i.e.\ solving Eqs.~(\ref{eq:f_ext_x}) - (\ref{eq:gamma_def})) in Fig.~\ref{fig:no_RR}. The initial parameters are $\gamma_0=10^4$, $x_1=0.1$ and $\zeta_0=0$. One may calculate the oscillation amplitude of $\zeta$ and $\dot{\gamma}$ are $8.84\times 10^{-6}$ and $1.33\times 10^{-5}$, respectively, according to Eqs.~(\ref{eq:zeta}), (\ref{eq:zeta1}) and (\ref{eq:gamma_dot0}). The amplitudes are confirmed by Fig.~\ref{fig:no_RR}. We also confirm the phase-locking Lorentz factor Eq.~(\ref{eq:gamma_z0}) by comparing the two subplots, i.e.\ Fig.~\ref{fig:no_RR} (a) shows a phase-locking case with $\gamma_w=\gamma_{z0}$ and Fig.~\ref{fig:no_RR} (b) shows a phase drifting case with $\gamma_w>\gamma_{z0}$.

\section{\label{sec:x1_gamma}Betatron Amplitude during Acceleration or Deceleration}
In this section we consider the effect of slowly varying $\gamma_0$. We only keep the first order variation. Within a time scale $t \sim \omega_\beta^{-1} \ll \gamma_0/\dot{\gamma_0}$, the change of $\gamma_0$ is
\begin{equation}
	\Delta \gamma_0 = \dot{\gamma_0} t.
\end{equation}
As a result, the amplitude and frequency of BO also change
\begin{eqnarray}
	\Delta x_1 &=& \dot{x_1} t,\\
	\Delta \omega_\beta &=& \frac{1}{2}\dot{\omega_\beta} t. \label{eq:Delta_omega_beta}
\end{eqnarray}
Note the factor $1/2$ in Eq.~(\ref{eq:Delta_omega_beta}) comes from the fact that a linear chirp contributes twice to the frequency shift. BO still follows Eq.~(\ref{eq:betatron_x}) but with slow-varying $x_1$ and $\omega_\beta$
\begin{equation}
    \begin{aligned}
	    x &= \left(x_1 + \Delta x_1\right) \sin\left(\omega_\beta + \Delta \omega_\beta\right) t\\
	      &= \left(x_1 + \dot{x_1} t\right) \sin\phi,
    \end{aligned}
\end{equation}
where $\phi \equiv \left(\omega_\beta + \frac{1}{2}\dot{\omega_\beta} t\right) t$. According to Eqs.~(\ref{eq:f_ext_x}) and (\ref{eq:EOM_no_rr}), we have
\begin{eqnarray}
	\dot{p}_x = -\frac{1}{2}x_1\left(1 + \frac{\dot{x_1}}{x_1} t\right) \sin\phi. \label{eq:f_ext_x_var_gamma}
\end{eqnarray} 

Meanwhile, we write down
\begin{equation}
    \begin{aligned}
        p_x &= \left(\gamma_0 + \Delta \gamma_0\right) \times \dot{x}\\
	        &= \gamma_0 x_1 \omega_\beta \left(1+\frac{\dot{\gamma_0}}{\gamma_0}t + \frac{\dot{x_1}}{x_1}t + \frac{\dot{\omega_\beta}}{\omega_\beta}t\right)\cos\phi + \gamma_0 \dot{x_1} \sin\phi,
    \end{aligned}
\end{equation}
where $\frac{\dot{\gamma_0}}{\gamma_0}t$, $\frac{\dot{x_1}}{x_1}t$ and $\frac{\dot{\omega_\beta}}{\omega_\beta}t$ are in the same order, and higher order terms are neglected. Then we take derivative one more time
\begin{equation}
    \begin{aligned}
        \dot{p_x} &= -\gamma_0 x_1 \omega_\beta^2 \left(1+\frac{\dot{\gamma_0}}{\gamma_0}t + \frac{\dot{x_1}}{x_1}t + 2\frac{\dot{\omega_\beta}}{\omega_\beta}t\right)\sin\phi + \gamma_0 x_1 \omega_\beta \left(\frac{\dot{\gamma_0}}{\gamma_0} + 2\frac{\dot{x_1}}{x_1} + \frac{\dot{\omega_\beta}}{\omega_\beta}\right)\cos\phi. \label{eq:px_dot_var_gamma}
    \end{aligned}
\end{equation}

We substitute left-hand-side of Eq.~(\ref{eq:f_ext_x_var_gamma}) by Eq.~(\ref{eq:px_dot_var_gamma}), the 0th order of the ``sin'' term retrieves Eq.~(\ref{eq:omega_beta}). The 1st order of the ``sin'' and ``cos'' terms construct two equations
\begin{eqnarray}
	\frac{\dot{\gamma_0}}{\gamma_0} + 2\frac{\dot{\omega_\beta}}{\omega_\beta} &=& 0,\\
	\frac{\dot{\gamma_0}}{\gamma_0} + 2\frac{\dot{x_1}}{x_1} + \frac{\dot{\omega_\beta}}{\omega_\beta} &=& 0,
\end{eqnarray}
which lead to
\begin{eqnarray}
	 \frac{\dot{\omega_\beta}}{\omega_\beta} &=& -\frac{1}{2}\frac{\dot{\gamma_0}}{\gamma_0}, \label{eq:omega_beta_dot_gamma_dot}\\
	\frac{\dot{x_1}}{x_1} &=& -\frac{1}{4}\frac{\dot{\gamma_0}}{\gamma_0}. \label{eq:x1_dot_gamma_dot}
\end{eqnarray}
By integral, we obtain the long-term dependencies of $\omega_\beta$ and $x_1$ on $\gamma_0$
\begin{eqnarray}
	 \omega_\beta &\propto& \gamma_0^{-\frac{1}{2}},\\
	 x_1 &\propto& \gamma_0^{-\frac{1}{4}}. \label{eq:x1_gamma}
\end{eqnarray}
The amplitude of $p_x$ oscillation is
\begin{eqnarray}
	 p_{x1} = x_1 \gamma_0 \omega_\beta = \frac{1}{\sqrt{2}}x_1 \gamma_0^{\frac{1}{2}}. \label{eq:px1}
\end{eqnarray}
As a result, the area encircled by the trajectory of the particle in the $x$-$p_x$ phase space,
\begin{eqnarray}
	 S = \pi x_1 p_{x1} = \frac{\pi}{\sqrt{2}}x_1^2 \gamma_0^{\frac{1}{2}} = \frac{\pi}{\sqrt{2}} A^2, \label{eq:S_def}
\end{eqnarray}
is a constant with varying $\gamma_0$, where $A$ is defined as the normalized BO amplitude
\begin{equation}
    A \equiv  x_1\gamma_0^{\frac{1}{4}}. \label{eq:A}
\end{equation}

\begin{figure}
    \centering
    \begin{overpic}[width=0.49\textwidth]{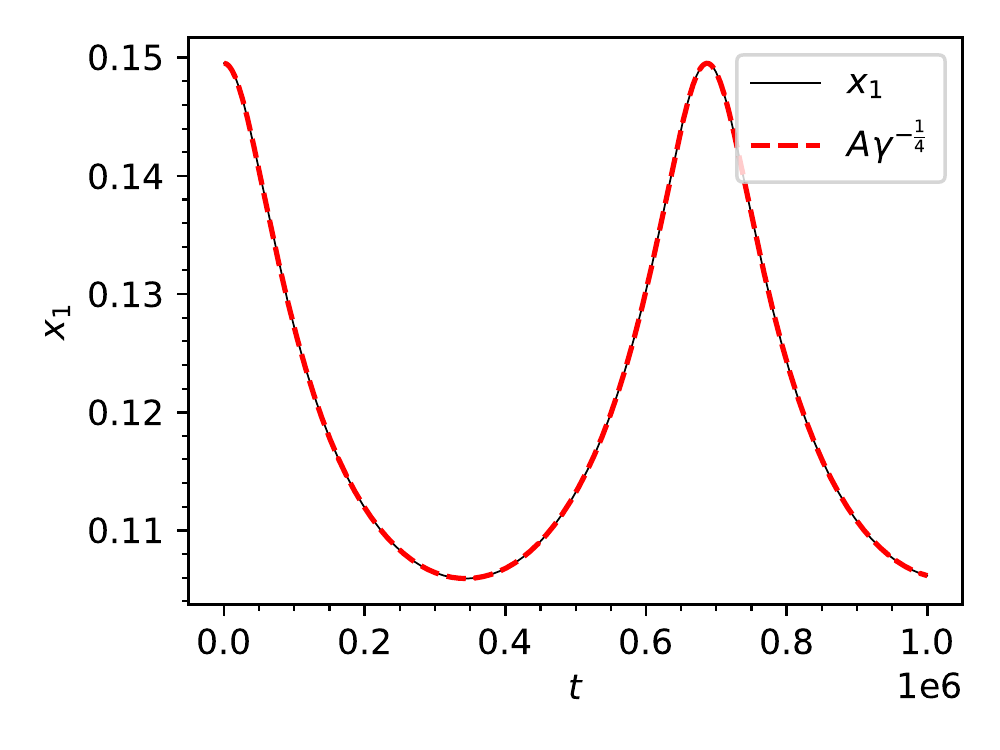}
        \put(23,20){(a)}
    \end{overpic}
    \begin{overpic}[width=0.49\textwidth]{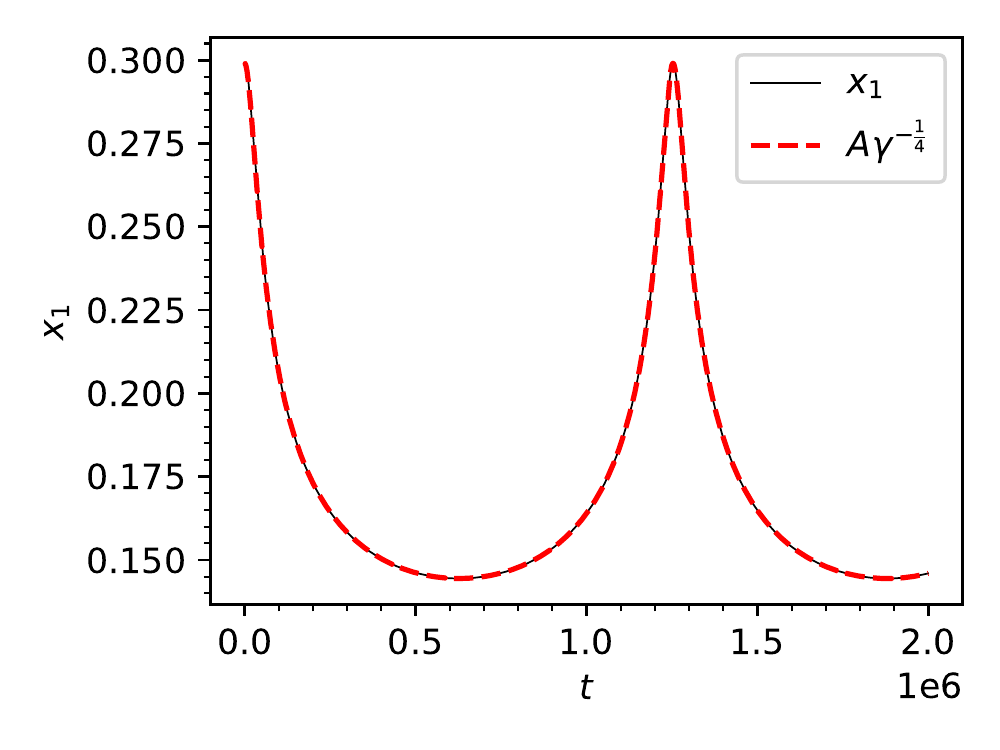}
        \put(25,20){(b)}
    \end{overpic}
    \caption{Numerical results of BO amplitude $x_1$ (black solid curves) and the scaling of $\gamma^{-1/4}$ (red dashed curves) during acceleration and deceleration processes. (a) The parameters at $t=0$ are $\gamma_0=2000$, $\zeta_0=0$, $\gamma_w=1000$ and $x_1=A\gamma_0^{-1/4}$ with $A=1$. (b) The parameters are similar to (a), but with $A=2$.}
    \label{fig:x1_gamma}
\end{figure}

We check the scaling Eq.~(\ref{eq:x1_gamma}) using PTracker. Two cases with significant acceleration and deceleration are shown in Fig.~\ref{fig:x1_gamma}. Because $\zeta_0=0$ and $\gamma_{z0}<\gamma_w$ at $t=0$, the electron firstly drifts to the acceleration phase, and $\gamma_0$ increases so that $\gamma_{z0}>\gamma_w$ at some point. Later it crosses the $\zeta=0$ point, enter the deceleration phase and $\gamma_0$ starts to decrease. The two numerical cases show exact scaling of $x_1 = A\gamma^{-1/4}$. 
Note we have replaced $\gamma_0$ by $\gamma$ here, because their difference is not distinguishable in these plots, and $\gamma$ can be obtained much more easily than $\gamma_0$ from numerical result.

\section{\label{sec:RR}Radiation Reaction Effects on the Betatron Oscillation}
The Lorentz-Abraham-Dirac equation in the same normalized units as in Sec.~\ref{sec:BO} is
\begin{eqnarray}
	F^{\rm rad}_\mu=\frac{2}{3}r_e\left[\frac{d^2p_\mu}{d\tau^2}-p_\mu\frac{dp_\nu}{d\tau}\frac{dp^\nu}{d\tau}\right] \label{eq:F_rad_mu}
\end{eqnarray}
with the metric $(-1, 1, 1, 1)$, where $\tau$ is the proper time ($\gamma d\tau=dt$), $r_e$ is the classical electron radius (also normalized to $k_p^{-1}$), $p^\mu=\left(\gamma,\ \vec{p}\right)$ is the four-momentum, and $F^{\rm rad}_\mu$ is the four-force of RR. The space component of Eq.~(\ref{eq:F_rad_mu}) is
\begin{eqnarray}
	\gamma\vec{f}^{\rm rad}=\frac{2}{3}r_e\left[\gamma\frac{d\gamma\dot{\vec{p}}}{dt}-\vec{p}\gamma^2\left(-\dot{\gamma}^2+|\dot{\vec{p}}|^2\right)\right].
\end{eqnarray}
The three-dimensional RR force $\vec{f}^{\rm rad}$ is decomposed to
\begin{eqnarray}
	\vec{f}^{\rm rad1}&=&\frac{2}{3}r_e\frac{d\gamma\dot{\vec{p}}}{dt},\label{eq:f_rad1}\\
	\vec{f}^{\rm rad2}&=&-\frac{2}{3}r_e\vec{p}\gamma\left(|\dot{\vec{p}}|^2-\dot{\gamma}^2\right).\label{eq:f_rad2}
\end{eqnarray}
With the RR force, the equation of motion Eq.~(\ref{eq:EOM_no_rr}) is modified as
\begin{eqnarray}
	\dot{\vec{p}} = \vec{f}^{\rm ext} + \vec{f}^{\rm rad1} + \vec{f}^{\rm rad2}. \label{eq:EOM_with_rr}
\end{eqnarray}

\subsection{Radiation reaction force of betatron oscillation}
We assume RR is perturbation to the electron BO which has the form discussed in Sec.~\ref{sec:BO},
\begin{eqnarray}
	\gamma &=& \gamma_0 + \frac{3}{32}x_1^2\cos2\omega_\beta t,\label{eq:gamma_expression3}\\
	p_x &=& x_1 \omega_\beta \gamma_0 \cos\omega_\beta t,\\
	p_z &=& \beta_{z0} \gamma_0 - \frac{x_1^2}{32}\cos2\omega_\beta t,\label{eq:pz1}\\
	\dot{\gamma} &=& \dot{\gamma_0} -\frac{3}{16} x_1^2 \omega_\beta \sin2\omega_\beta t, \label{eq:gamma_dot1}\\
	\dot{p_x} &=& -\frac{1}{2}x_1 \sin\omega_\beta t,\\
	\dot{p_z} &=& -\frac{1}{2}\zeta_0 + \frac{1}{16}x_1^2 \omega_\beta \sin2\omega_\beta t. \label{eq:pz_dot1}
\end{eqnarray}
Note Eq.~(\ref{eq:gamma0}), as long as $|\dot{\gamma_0}| \sim |\zeta_0| \lesssim 1 \ll \omega_\beta^{-1}$, we have
\begin{eqnarray}
    |\dot{\vec{p}}|^2-\dot{\gamma}^2 \approx \dot{p_x}^2. \label{eq:p_dot2_m_gamma_dot2}
\end{eqnarray} 
According to Eqs.~(\ref{eq:f_rad1}) and (\ref{eq:f_rad2}), the RR force terms are 
\begin{eqnarray}
	f^{\rm rad1}_x &=& -\frac{1}{3}r_e x_1 \gamma_0 \omega_\beta \cos\omega_\beta t = -\frac{1}{3}r_e p_x, \label{eq:f_rad1_x}\\
	f^{\rm rad1}_z &=& \frac{2}{3}r_e \dot{\gamma_0}^2 + \frac{1}{12}r_e x_1^2 \gamma_0 \omega_\beta^2 \cos2\omega_\beta t, \label{eq:f_rad1_z}\\
	f_x^{\rm rad2} &=& -\frac{1}{12}r_e x_1^2 \gamma_0 p_x\left(1 - \cos2\omega_\beta t\right), \label{eq:f_rad2_x}\\ 
	f_z^{\rm rad2} &=& -\frac{1}{12}r_e x_1^2 \gamma_0^2 \left(1 - \cos2\omega_\beta t\right). \label{eq:f_rad2_z}
\end{eqnarray}
Use Eqs.~(\ref{eq:x1_gamma}) and (\ref{eq:A}), we have $x_1^2\gamma_0 \gg 1$ and conclude $\left|f_x^{\rm rad2}\right| \gg \left|f_x^{\rm rad1}\right|$. Use Eq.~(\ref{eq:omega_beta}) and $|\dot{\gamma_0}| \lesssim 1$, we conclude $\left|f_z^{\rm rad2}\right| \gg \left|f_z^{\rm rad1}\right|$. We also have
\begin{equation}
    \frac{f_x^{\rm rad2}}{f_z^{\rm rad2}} = \beta_x = x_1\omega_\beta \cos\omega_\beta t \propto x_1 \gamma_0^{-\frac{1}{2}} \propto \gamma_0^{-\frac{3}{4}}. \label{eq:f_rad_x_over_f_rad_z}
\end{equation}

\subsection{\label{sec:pahselock_RR}Perturbation correction and phase-locking with longitudinal forces}
In Sec.~\ref{sec:BO} we have obtained the phase-locking Lorentz factor $\gamma_{z0}$ in Eq.~(\ref{eq:gamma_z0}). With RR, extra longitudinal deceleration force exists. Define the phase-locking $\zeta_0$ to be
\begin{equation}
    \zeta_{0l} \equiv -\frac{1}{6}r_e x_1^2 \gamma_0^2, \label{eq:zeta0l}
\end{equation}
the summation of the longitudinal external and RR forces is written by using Eqs.~(\ref{eq:f_ext_z}), (\ref{eq:zeta}), (\ref{eq:zeta1}) and (\ref{eq:f_rad2_z})
\begin{equation}
    \begin{aligned}
        f_z &= -\frac{\zeta_0-\zeta_{0l}}{2} + \frac{1}{16} x_1^2 \omega_\beta \sin2\omega_\beta t + \frac{1}{12}r_e x_1^2 \gamma_0^2\cos2\omega_\beta t. \label{eq:fz}
    \end{aligned}
\end{equation}
If $\zeta_0 = \zeta_{0l}$, the longitudinal momentum gain averaged in the $\omega_\beta^{-1}$ time scale is zero. If $\gamma_w = \gamma_{z0}$ is also satisfied, the phase-locking will occur and there will not be net acceleration or deceleration. The 2nd term, coming from BO, is already included in Eqs.~(\ref{eq:gamma_dot1}) and (\ref{eq:pz_dot1}). In cases the 3rd term, coming from RR, is comparable or more significant than the 2nd term, i.e.\ $r_e \gamma_0^{{5}/{2}}\gtrsim 1$, Eqs.~(\ref{eq:gamma_expression3}), (\ref{eq:pz1}), (\ref{eq:pz_dot1}) and (\ref{eq:gamma_dot1}) should be modified
\begin{eqnarray}
    \gamma &=& \gamma_0 + \frac{3}{32}x_1^2\cos2\omega_\beta t + \frac{1}{24}r_e x_1^2\frac{\gamma_0^2}{\omega_\beta}\sin2\omega_\beta t, \label{eq:gamma_expression4}\\
	p_z &=& \beta_{z0} \gamma_0 - \frac{x_1^2}{32}\cos2\omega_\beta t + \frac{1}{24}r_e x_1^2\frac{\gamma_0^2}{\omega_\beta}\sin2\omega_\beta t, \label{eq:pz2}\\
    \dot{\gamma} &=& \dot{\gamma_0} -\frac{3}{16} x_1^2 \omega_\beta \sin2\omega_\beta t + \frac{1}{12}r_e x_1^2 \gamma_0^2\cos2\omega_\beta t, \label{eq:gamma_dot_RR}\\
    \dot{p_z} &=& -\frac{\zeta_0-\zeta_{0l}}{2} + \frac{1}{16}x_1^2 \omega_\beta \sin2\omega_\beta t + \frac{1}{12}r_e x_1^2\gamma_0^2\cos2\omega_\beta t,\label{eq:pz_dot_RR}
\end{eqnarray}
where the $\gamma_0$ definition Eq.~(\ref{eq:gamma0}) is modified as
\begin{eqnarray}
	\left.\gamma_0\right|_{t_0}^{t_1} = - \frac{1}{2} \int_{t_0}^{t_1} \left(\zeta_0 - \zeta_{0l}\right) \beta_{z0} dt.
\end{eqnarray}
In the regime $r_e \gamma_0 \ll 1$, Eq.~(\ref{eq:p_dot2_m_gamma_dot2}) still holds, and in Eqs.~(\ref{eq:f_rad1_x}) - (\ref{eq:f_rad2_z}), only Eq.~(\ref{eq:f_rad1_z}) has to be modified
\begin{eqnarray}
	f^{\rm rad1}_z &=& \frac{2}{3}r_e \dot{\gamma_0}^2 + \frac{1}{12}r_e x_1^2 \gamma_0 \omega_\beta^2 \cos2\omega_\beta t - \frac{1}{9}r_e^2 x_1^2 \gamma_0^3 \omega_\beta \sin2\omega_\beta t.
\end{eqnarray}
Note $\left|f_z^{\rm rad2}\right| \gg \left|f_z^{\rm rad1}\right|$ still holds, the above discussion is self-consistent. One may compare our Eqs.~(\ref{eq:gamma_dot_RR}) and (\ref{eq:pz_dot_RR}) with Eqs.~(30) and (32) of Ref.~\cite{AHDengPRAB2012} by taking $K^2=k_p^2/2$ and find that our model keeps more details of the BO, while they have omitted the difference between $\dot{\gamma}$ and $\dot{p_z}$. These details, as shown in Sec.~\ref{sec:BOdamp}, are important for the long term RR effect.

We also estimate the maximum $\gamma_0$ achievable in PWA, due to the limited size of the blowout wakefield structure, by setting $\zeta_{0l}\sim -1$ in Eq.~(\ref{eq:zeta0l}) and using Eq.~(\ref{eq:S_def}):
\begin{equation}
    \gamma_{0\max} \sim \left(\frac{3\sqrt{2}\pi}{r_e S}\right)^{\frac{2}{3}} = \left(\frac{6}{r_e A^2}\right)^{\frac{2}{3}}. \label{eq:gamma0max}
\end{equation}
One should note that this maximum energy is not due to dephasing or pump-depletion effects. It is the power of radiation loss approximately equals to the maximum power of acceleration in the wakefield.

The numerical solution for a case of phase-locking by PTracker with RR, i.e.\ Eq.~(\ref{eq:EOM_no_rr}) be replaced by Eq.~(\ref{eq:EOM_with_rr}), is shown in Fig.~\ref{fig:RR} (a). With the $t=0$ parameters $\gamma_0=10^5$, $x_1=\gamma_0^{-1/4}$, $\gamma_w=\gamma_{z0}$ according to Eq.~(\ref{eq:gamma_z0}) and $\zeta_0=\zeta_{0l}$ according to Eq.~(\ref{eq:zeta0l}), the numerical result shows a phase-locking with an oscillation amplitude $\zeta_1 = 8.84\times 10^{-7}$ which confirms Eq.~(\ref{eq:zeta1}).

\subsection{\label{sec:BOdamp}Damping of betatron oscillation due to radiation fraction}
The potential and the kinetic energy in the transverse direction transform to each other due to BO, with the maximum potential energy being
\begin{equation}
    U = \frac{x_1^2}{4} \label{eq:U}.
\end{equation}
Meanwhile, this potential energy is slowly lost due to the work of dissipation force averaged in the $\omega_\beta^{-1}$ time scale. The averaged power of the transverse force is
\begin{equation}
    \begin{aligned}
        \left<\beta_x f_x\right> &= \left<\beta_x \left(f^{\rm ext}_x + f^{\rm rad1}_x+f^{\rm rad2}_x\right)\right> \\
        &= 0 -\frac{1}{3}r_e\left<\frac{p_x^2}{\gamma}\right> - \frac{1}{12}r_e x_1^2 \gamma_0 \left<\left(1-\cos2\omega_\beta t\right)\frac{p_x^2}{\gamma}\right> \\
        &= -\frac{1}{12}r_e x_1^2 \left(1+\frac{1}{8}x_1^2\gamma_0\right)\\
        &\approx -\frac{1}{96}r_e x_1^4 \gamma_0,
    \end{aligned}
\end{equation}
and the averaged power of the longitudinal forces by using Eqs.~(\ref{eq:fz}), (\ref{eq:gamma_expression4}) and (\ref{eq:pz2}) is
\begin{eqnarray}
    \left<\beta_z f_z\right> &=& \left< \left(\beta_{z0}- \frac{1}{8}\frac{x_1^2}{\gamma_0}\cos2\omega_\beta t\right) \left(-\frac{\zeta_0-\zeta_{0l}}{2} + \frac{1}{16} x_1^2 \omega_\beta \sin2\omega_\beta t + \frac{1}{12}r_e x_1^2 \gamma_0^2\cos2\omega_\beta t\right)\right> \nonumber \\
        &=& -\frac{1}{2} \left(\zeta_0-\zeta_{0l}\right) \beta_{z0} -\frac{1}{192}r_e x_1^4 \gamma_0, \label{eq:beta_z_f_z_ave}
\end{eqnarray}
where the angle brackets stands for averaging in the $\omega_p^{-1}$ time scale. Note the 1st term in the right-hand-side of Eq.~(\ref{eq:beta_z_f_z_ave}) contributes to the longitudinal acceleration or deceleration, while the 2nd term, coming from the coupling of the oscillation terms of $\beta_z$ and $f^{\rm rad2}_z$, contributes to the transverse damping of BO. Thus
\begin{equation}
    \begin{aligned}
        \dot{U} = \left<\beta_x f_x\right> + \left.\left<\beta_z f_z\right>\right|_{\rm 2nd\ term} = -\frac{1}{64}r_e x_1^4 \gamma_0,
    \end{aligned}
\end{equation}
which can be applied to Eqs.~(\ref{eq:U}) for the damping rate of the BO amplitude due to RR
\begin{equation}
    \left.\frac{\dot{x_1}}{x_1}\right|^{\rm rad} = -\frac{1}{32}r_e \gamma_0 x_1^2. \label{eq:x1_damp_rad}
\end{equation}
Note Eq.~(\ref{eq:x1_dot_gamma_dot}), the total damping rate of $x_1$ is
\begin{equation}
    \frac{\dot{x_1}}{x_1}= -\frac{1}{32}r_e \gamma_0 x_1^2 -\frac{1}{4}\frac{\dot{\gamma_0}}{\gamma_0}. \label{eq:x1_damp}
\end{equation}
Use Eqs.~(\ref{eq:S_def}) and (\ref{eq:x1_damp}), we find the damping rate of the area encircled by the trajectory of the particle in the $x$-$p_x$ phase space
\begin{equation}
    \frac{\dot{S}}{S} = -\frac{1}{16}r_e \gamma_0 x_1^2 = -\frac{1}{8\sqrt{2}\pi}r_e \gamma_0^{\frac{1}{2}} S, \label{eq:S_dot}
\end{equation}
or in the integrated form
\begin{equation}
    \left.\frac{1}{S}\right|_{t_0}^{t_1} = \frac{1}{8\sqrt{2}\pi}r_e \int_{t_0}^{t_1} \gamma_0^{\frac{1}{2}} dt. \label{eq:S_damp_int}
\end{equation}
In the time scale that $\gamma_0$ does not change significantly, it can be simplified as
\begin{equation}
    \left.\frac{1}{S}\right|_{t_0}^{t_1} = \frac{1}{8\sqrt{2}\pi}r_e \gamma_0^{\frac{1}{2}} \left(t_1 - t_0\right). \label{eq:S_damp_int_const_gamma}
\end{equation}
Thus the length that $S$ reduces by a half is
\begin{equation}
    L_S = \frac{8\sqrt{2}\pi}{r_e \sqrt{\gamma_0} S} = \frac{16}{r_e \sqrt{\gamma_0} A^2}, \label{eq:LS}
\end{equation}
or in the unnormalized form (note $S$ and $A$ remain in normalized units)
\begin{equation}
    k_p L_S = \frac{8\sqrt{2}\pi}{k_p r_e \sqrt{\gamma_0} S} = \frac{16}{k_p r_e \sqrt{\gamma_0} A^2}.
\end{equation}

\begin{figure}
    \centering
    \begin{overpic}[width=0.49\textwidth]{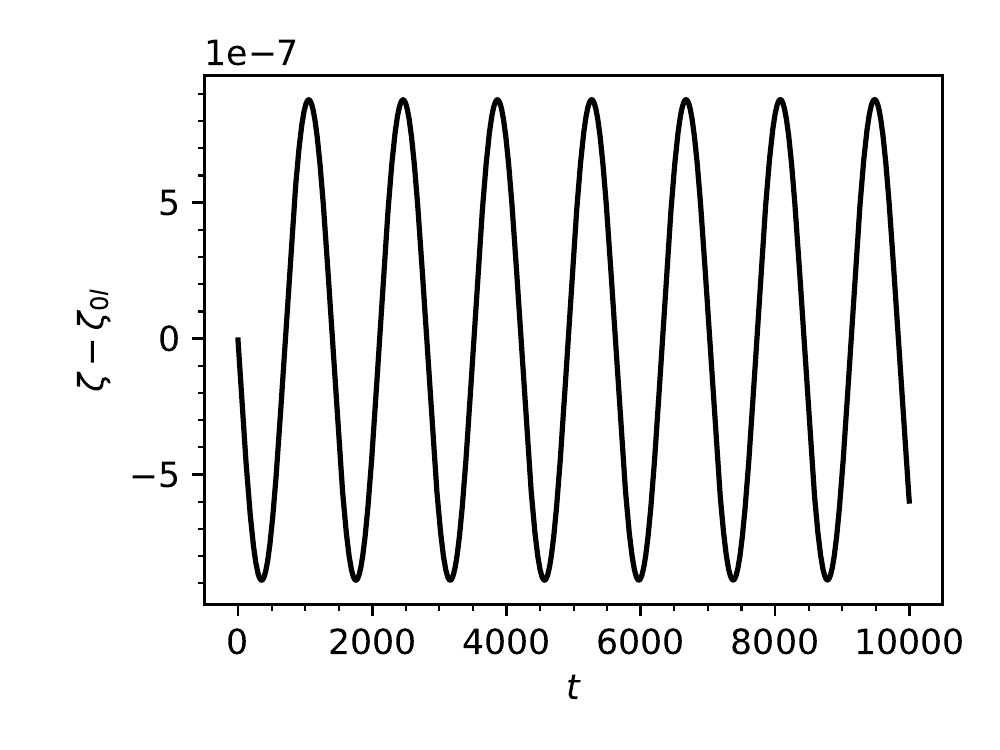}
        \put(23,60){(a)}
    \end{overpic}
    \begin{overpic}[width=0.49\textwidth]{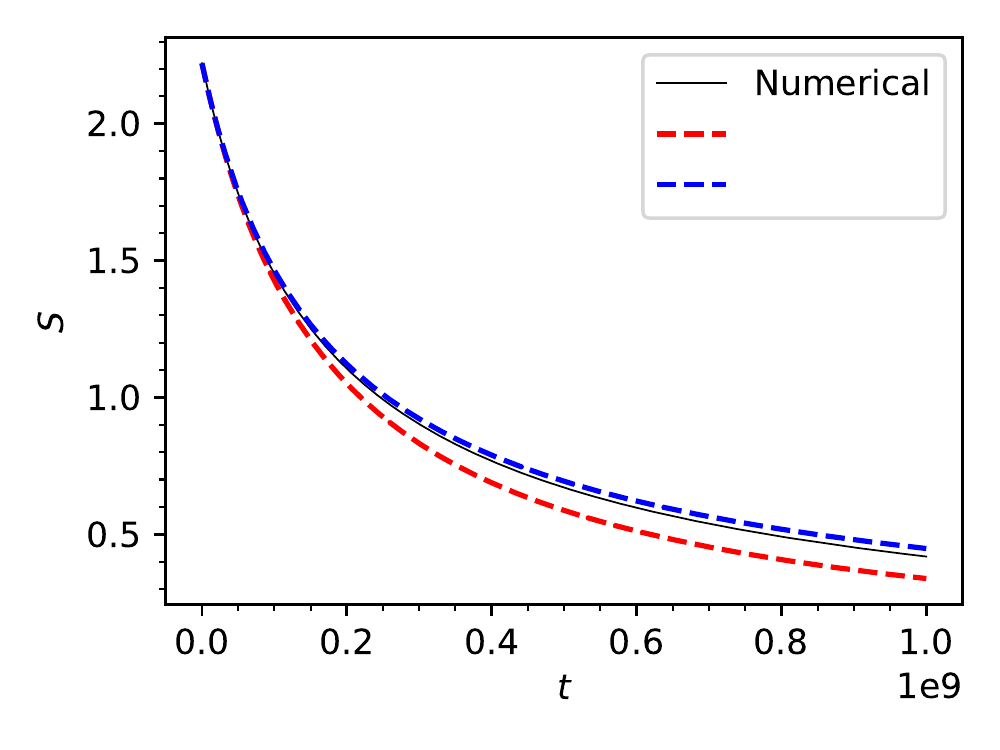}
        \put(75,60){Eq.~(\ref{eq:S_damp_int_const_gamma})}
        \put(75,55){Eq.~(\ref{eq:S_damp_int})}
        \put(20,20){(b)}
    \end{overpic}
    \caption{Numerical results using PTracker with RR turned on. The length normalization unit is $k_p^{-1}=10^{-5}\ \rm m$, thus the normalized classical electron radius is $r_e=2.82\times 10^{10}$. The parameters at $t=0$ are $\gamma_0=10^5$, $x_1=\gamma_0^{-1/4}$, $\gamma_w=\gamma_{z0}$ and $\zeta_0=\zeta_{0l}$. (a) The oscillation of $\zeta$ at an early stage. (b) The damping of $S$, defined by Eq.~(\ref{eq:S_def}), in a long time scale. Numerical solution of $S$ (black solid curve) is compared to Eqs.~(\ref{eq:S_damp_int_const_gamma}) (red dashed curve) and (\ref{eq:S_damp_int}) with numerical integral (blue dashed curve).}
    \label{fig:RR}
\end{figure}

For an electron beam which has an distribution of $S$, the damping rates are different for different $S$. The reduction rate of the normalized emittance $\epsilon_n \approx S/k_p$ is estimated by Eq.~(\ref{eq:S_dot}), but with $S$ replaced by the largest $S$ among the electrons. Thus, we obtain the engineering formula of the normalized emittance reduction length
\begin{equation}
    L_{\epsilon_n} [{\rm m}] = \frac{1.89\times 10^6}{\sqrt{\gamma_0 (n_p [{10^{18}\ \rm cm^{-3}}])^3} \cdot \epsilon_n [{\rm mm\cdot mrad}]}.
\end{equation}

The numerical solution for the $S$ damping in a long time scale ($\Delta t = 10^9 > L_S = 1.8 \times 10^8$) with the same parameters as Fig.~\ref{fig:RR} (a) is shown in Fig.~\ref{fig:RR} (b). Eq.~(\ref{eq:S_damp_int_const_gamma}) predicts the numerical result perfectly until at about $t=10^8$, because $\gamma_0$ starts to change due to the breaking of phase-locking condition while $x_1$ is damped. Eq.~(\ref{eq:S_damp_int}) with numerical integral matches the numerical result even better at an extremely long time scale, but the integral of $\gamma_0^{1/2}$ is replaced by the integral of $\gamma^{1/2}$ because $\gamma$ can be much more easily obtained than $\gamma_0$ from numerical result.

\section{\label{sec:CQL}From Classical to Quantum Radiation Domain}

The correction by nonlinear quantum electrodynamics (QED) has been studied for high-intensity laser-electron interactions~\cite{Ritus1985, ADiPiazzaRMP2012, XLZhuNJP2018}. The radiation power with nonlinear QED correction is 
\begin{align}
    P & = q(\chi) P_\mathrm{classical}, \label{eq:QED_P}
\end{align}
where $P_\mathrm{classical}$ is the classical radiation power, and $q(\chi)$ is the QED correction factor~\cite{Sokolov2010, Sokolov1986, SETO2021}. $q(\chi) \approx 1$ with $\chi \ll 1$ is the classical regime of the radiation process, while $q(\chi) < 1$ with $\chi \gtrsim 1$ is the QED regime. We calculate the significance of this correction in our model.

RR is effective when a background force is strong enough and perpendicular to the electron momentum in the relativistic regime. This perpendicular direction is primarily the $x$-direction in our case. We write down the effective vector potential amplitude of the wakefield
\begin{align}
    \mathcal{W}_0 = \mathcal{O}(p_{x1}) \sim x_1 \gamma^{\frac{1}{2}},
\end{align}
which is an analogy of the normalized laser vector potential amplitude $a_0$ in laser-electron interactions. When we consider a radiation process to be nonlinear Compton scattering by a laser pulse, the formation phase (or the formation length) $a_0^{-1}$ is important for the locally constant field approximation (LCFA) in the semi-classical model~\cite{Ritus1985, ADiPiazzaRMP2012, XLZhuNJP2018, Sokolov2010, Nikishov1964_01, Nikishov:1964_02, Sokolov2010, SETO2021}. Namely, a single photon is emitted instantaneously in a short phase interval $a_0^{-1}$ at high intensities. By analogy, the formation phase in the wakefield is $\mathcal{O}(\mathcal{W}_0^{-1})$ which is short enough if $\gamma \gg 1$. Thus, the LCFA of a radiation process can be employed for the high-energy electron case. Also, the quantum correction $q(\chi)$ can be applied, with the dimensionless quantum parameter
\begin{equation}
    \begin{aligned}
        \chi = \frac{r_e}{\alpha}\left| F^\mathrm{ext}_\mu \right| 
        & = \frac{r_e}{\alpha} \sqrt{ - \left(\gamma \vec{f}^\mathrm{ext} \cdot \vec{\beta}\right)^2 + \left|\gamma \vec{f}^\mathrm{ext}\right|^2  } \\
        & = \frac{r_e}{2 \alpha} \gamma \sqrt{ x^2 (1-\beta_x^2) + \zeta^2 (1-\beta_z^2)  } \\
        & \approx \frac{r_e}{2 \alpha} |x| \gamma,
    \end{aligned}
\end{equation}
where $\alpha$ is the fine structure constant, $F^\mathrm{ext}_\mu = (-\gamma \vec{f}^\mathrm{ext} \cdot \vec{\beta},\ \gamma \vec{f}^\mathrm{ext})$ is the four-force of the wakefield, $\beta_x \sim x_1 \gamma^{-1/2} \ll 1$ and $1-\beta_z^2 \approx \gamma_{z0}^{-2} \sim x_1^2/\gamma \ll x^2$. Note $r_e$ is normalized to $k_p^{-1}$ in the expression. The maximum $\chi$ in one BO cycle is
\begin{equation}
    \chi_{\max} \approx \frac{r_e}{2\alpha} x_1 \gamma = \frac{r_e}{2\alpha} A \gamma^{\frac{3}{4}} = \frac{r_e}{\alpha} \sqrt{\frac{S}{\pi}} \left( \frac{\gamma}{2} \right)^{\frac{3}{4}}. \label{eq:eta_max}
\end{equation}
For $\chi_{\max}\ll 1$, RR is classical, while for $\chi_{\max}\gtrsim 1$, quantum correction may appear in Eq.~(\ref{eq:QED_P}). Commonly, PWA with internal injections has $A\sim 1$. Thus for $k_p\sim 10^5\ \rm m^{-1}$ ($r_e\sim 10^{-10}$ as a consequence), RR is classical if $\gamma\ll 10^{10}$.

\section{Summary and Discussion}
We have developed a classical model of the electron betatron oscillation with radiation reaction (RR) in a plasma wakefield accelerator. We found a phase-locking condition $\gamma_w = \gamma_{z0}$ and $\zeta_0 = \zeta_{0l}$ using Eqs.~(\ref{eq:gamma_z0}) and (\ref{eq:zeta0l}), under which the electron does not gain or loss energy. The maximum $\gamma_0$ achievable in a plasma wakefield accelerator due to RR is estimated by Eq.~(\ref{eq:gamma0max}). We also found the damping rate of $S$, defined as the area encircled by the electron trajectory in the $x$-$p_x$ phase space, in Eq.~(\ref{eq:S_dot}) and the length that $S$ reduces by a half in Eq.~(\ref{eq:LS}). The quantum parameter characterizing the classical and quantum radiation domain is given by Eq.~(\ref{eq:eta_max}).

Some examples with different plasma densities $n_p$, initial $S$ and $\gamma_0$ are listed in Tab.~\ref{tab:example}. We can see that the reduction of $S$ (and thus the transverse cooling) is positively related to 1) the plasma density, 2) the Lorentz factor of the electron, 3) the initial betatron amplitude and 4) the total length of the plasma wakefield accelerator. We also note that the quantum parameter $\chi_{\max}$ is sufficiently smaller than unity for all these cases.
\begin{table}
    \centering
    \begin{tabular}{c|c|c|c|c|c|c|c}
        \hline
        Case No. & $n_p$ [cm$^{-3}$] & $k_p$ [m$^{-1}$] & $S$ & $\gamma_{0\max}$ & $\gamma_0$ & $L_S$ [m] & $\chi_{\max}$ \\
        \hline
        1 & \multirow{4}{20pt}{$10^{18}$} & \multirow{4}{50pt}{$1.88\times 10^5$} & \multirow{2}{20pt}{2} & \multirow{2}{40pt}{$5.4\times10^6$} & $1\times10^5$ & 563.2 & $1.9\times 10^{-4}$ \\ \cline{1-1} \cline{6-8}
        2 &  &  &  &  & $5\times10^6$ & 80.0 & $3.6\times 10^{-3}$ \\ \cline{1-1} \cline{4-8}
        3 &  &  & \multirow{2}{20pt}{8} & \multirow{2}{40pt}{$2.1\times 10^6$} & $1\times10^5$ & 140.8 & $3.9\times 10^{-4}$ \\ \cline{1-1} \cline{6-8}
        4 &     &  &  &  & $2\times10^6$ & 31.5 & $3.7\times 10^{-3}$ \\
        \hline
        5 & \multirow{4}{20pt}{$10^{17}$} & \multirow{4}{50pt}{$5.95\times 10^4$} & \multirow{2}{20pt}{2} & \multirow{2}{40pt}{$1.2\times10^7$} & $1\times10^5$ & 5632 & $6.1\times 10^{-5}$ \\ \cline{1-1} \cline{6-8}
        6 &  &  &  &  & $1\times10^7$ & 563.2 & $1.9\times 10^{-3}$ \\ \cline{1-1} \cline{4-8}
        7 &  &  & \multirow{2}{20pt}{8} & \multirow{2}{40pt}{$4.6\times 10^6$} & $1\times10^5$ & 1408 & $1.2\times 10^{-4}$ \\ \cline{1-1} \cline{6-8}
        8 &  &  &  &  & $4\times10^6$ & 222.6 & $2.0\times 10^{-3}$ \\
        \hline
    \end{tabular}
    \caption{Some examples of $\gamma_{0\max}$, $L_S$ and $\chi_{\max}$ with varying $n_p$, $S$ and $\gamma_0$.}
    \label{tab:example}
\end{table}
 
In order to decrease $L_S$ without an extremely large $\gamma_0$, we can either increase $n_p$ or $S$. Although an internally injected electron beam has $S\sim 2$, a large $S$ is achievable by an external injection. In the two-stage scheme where the first stage with a smaller $n_p$ is used for acceleration and the second stage with a larger $n_p$ is used for radiation~\cite{XLZhuSA2020}, both $n_p$ and $S$ are large for the second stage. Thus RR takes effect in a much shorter distance in this case. However, $\chi$ may be not negligible for large $S$ values and quantum RR should be taken into consideration.

\section*{Acknowledgements}
MZ acknowledges fruitful discussions with X.-L.~Zhu. This work is supported by Research Foundation of Institute of High Energy Physics, Chinese Academy of Sciences (Grant No.\ E05153U1 and No.\ E15453U2), and the Romanian Ministry of Research and Innovation through the core program of the Ministry of Research PN 19 06 01 05. 
\bibliography{main}
\end{document}